\documentclass[useAMS,usenatbib,usegraphicx]{mn2e}
\usepackage{amsfonts}
\usepackage{amsmath}
\usepackage{amssymb}
\usepackage{cleveref}			% Used for clever cross referencing
\usepackage{longtable}
\usepackage{lscape}
\usepackage{pifont}
\usepackage{rotating}
\usepackage{stfloats}
\usepackage{tabularx}
\usepackage{tikz}				% Used for custom vector graphics.
\usepackage{times}
\usepackage{url}
\usepackage{graphicx}
\usepackage{caption}
\usepackage{subcaption}
\def\arcdeg{\hbox{$^\circ$}}
\def\arcmin{\hbox{$^\prime$}}
\def\arcsec{\hbox{$^{\prime\prime}$}}

\def\HII{H\,\textsc{ii}}
\def\ha{H$\alpha$}
\def\hb{H$\beta$}

\def\NII{[N\,\textsc{ii}]}

\def\OI{[O\,\textsc{i}]}

\def\OIII{[O\,\textsc{iii}]}

\def\SII{[S\,\textsc{ii}]}

\def\scos{SuperCOSMOS}

\def\p0{\phantom{0}}

\def\lessim{\raise-.5ex\hbox{$\buildrel<\over{\scriptstyle\mathtt{\approx}}$}}
\def\grtsim{\raise-.5ex\hbox{$\buildrel>\over{\scriptstyle\mathtt{\approx}}$}}

\title[Four New Planetary Nebulae Toward the SMC]{Four New Planetary Nebulae Toward the Small Magellanic Cloud}
\author[Dra\v{s}kovi\'{c} et al.]{Danica Dra\v{s}kovi\'{c}$^{1}$\thanks{E-mail.danica.draskovic@mq.edu.au}, Quentin A. Parker$^{1,2,3}$, 
Warren A. Reid$^{1}$ and Milorad Stupar$^{1}$\\
$^{1}$Department of Physics and Astronomy, Macquarie University, Sydney, NSW 2109, Australia\\
$^{2}$Australian Astronomical Observatory, PO Box 915, North Ryde, NSW 1670, Australia\\
$^{3}$Department of Physics, Chong Yeut Ming Physics building, The University of Hong Kong, Pokfulam, Hong Kong}
\begin{document}

\date{Accepted $<date>$. Received $<date>$; in original form $<date>$}

\pagerange{\pageref{firstpage}--\pageref{lastpage}} \pubyear{2014}

\maketitle

\label{firstpage}

\begin{abstract}
We present four new planetary nebulae (PNe) discovered in the Small Magellanic Cloud (SMC) from deep UK Schmidt telescope (UKST) narrow band \ha{} and broad-band short-red ``SR" continuum images and confirmed spectroscopically. All new PNe show strong \NII{}/\ha{} ratios in their spectra. We describe and detail the process of PN candidate selection based on wide-field multi-wavelength imaging of the SMC and our subsequent spectroscopic confirmation and classification. We carefully reviewed archived information and available imagery for previous SMC PN detections and various other types of emission objects in the SMC as a training set to help identify new PN candidates. These 4 preliminary discoveries provide a 4\% increase to the previously known SMC PN population of $\sim$100. Once spectroscopic follow-up of all our newly identified SMC PN candidates is complete, we expect to increase the total number of known SMC PNe by up to 50\%. This will permit a significant improvement to determination of the SMC PN luminosity function and enable further insights into the chemical evolution and kinematics of the SMC PN population.
\end{abstract}

\begin{keywords}
planetary nebulae: general -- Magellanic Clouds.
\end{keywords}

\section{Introduction}

{Planetary nebulae (PNe) represent an important but brief ($\sim$25,000~year), late stage of stellar evolution experienced by most low- and intermediate-mass stars (${\approx}1\mbox{--}8\,M_{\sun}$). They play a crucial role in understanding various aspects of late stellar evolution, such as mass loss  \citep{1995PhR...250....2I} and the subsequent interstellar enrichment by the products of nucleosynthesis like Oxygen, Nitrogen and dust. Their ionised gas shells exhibit numerous, strong emission lines that are excellent laboratories for understanding plasma physics. PNe are visible to great distances due to these strong lines that permit determination of nebula size, age and expansion velocity. The luminosity, temperature and mass of their central stars (CSPN) can also be estimated \citep{1993IAUS..155...65P, 2000ApJ...543..889G}, as can the chemical composition of the ejected gas \citep{2005ApJ...622..294S}. Determining PNe radial velocities also allows us to trace the kinematic properties of their host galaxies. Their complex morphologies provide clues to their formation, evolution and mass-loss processes.  This shaping is thought to result from a variety of mechanisms, including radiation-driven stellar winds \citep{2002ARA&A..40..439B},  CSPN binarity \citep{2009PASP..121..316D, 2009AAp...496..813M, 2009AAp...505..249M}, magnetic fields \citep{2001Natur.409..485B} and sub-stellar PN companions \citep{2011PASP..123..402D}.

Knowing PNe distances is key since important physical quantities like total nebular mass, luminosity and evolutionary states of the CSPN, as well as the physical extent of inner shells and outer haloes are distance dependant \citep{1999AJ....118..488C}. However, study of Galactic PNe suffers from severe problems with distance determinations \citep{1994A&AS..108..485V, 2001A&A...374..599B, 2015MNRAS.449.2980S} though significant recent progress  has been made \citep{2015arXiv150401534F}. Strongly varying dust obscuration across the Galactic plane and the  presence of  significant non-PN mimics have also biased studies based on previous Galactic PN catalogues, e.g. see \citet{2010PASA...27..129F}. These distance and extinction problems are largely  circumvented by studying PNe in nearby galaxies with well determined distances and extinctions such as the Large and Small Magellanic Clouds (LMC and SMC respectively). They are sufficiently close for detailed individual scrutiny by ground and space-based telescopes such as the \textit{Hubble Space Telescope} (HST)\citep{2003ApJ...596..997S, 2006ApJS..167..201S}. Studies between PNe residing in galaxies with different morphologies, metallicities, star forming history and chemical evolution \citep{2006IAUS..234....9M} can also be valuably  performed and would benefit from having statistically significant samples for comparison.

Population synthesis shows that a PN population, scaled to completeness limits, correlates with the visual magnitude of the host galaxy \citep{2003A&A...407...51M}. This allows the PN 
specific luminosity rate to be estimated for various local group galaxies \citep{1980ApJS...42....1J}. Based on the theoretical luminosity-specific PN density, \citet{2006pnbm.conf..355B} and \citet{2012IAUS..283..227R} used best determinations of bolometric luminosities and the current number of PNe known in local group galaxies to provide an estimate of the number of Local Group PNe that may yet be found. 

The LMC currently hosts $\sim$740 confirmed PNe \citep{2006MNRAS.365..401R, 2006MNRAS.373..521R, 2013MNRAS.436..604R, 2014MNRAS.438.2642R} compared to the 800--900 expected from population synthesis, while the SMC only has $\sim$100 comprising only $\sim$46\% of the expected total of $\sim$216 \citep{2002AJ....123..269J} for a complete survey (8 mag down the PNLF). More locally the Sagittarius Dwarf Galaxy (SagDG)  \citep{2006MNRAS.369..875Z} only has 4 currently known PNe. We have identified several further SagDG PN candidates using similar techniques to those we have described here (Dra\v{s}kovi\'{c} et al., in preparation).

Here we are concentrating on the SMC whose accurately determined distance of 61$\pm1$~kpc  \citep{2005MNRAS.357..304H} allows derivation of true PNe physical parameters from 
their optical spectra and other observed properties. This makes them powerful tools for studies of late stage stellar evolution in this extragalactic environment of different metallicity to both our Galaxy and the LMC. Line of sight dimming towards the SMC by intervening gas and dust is also generally low and uniform, enabling absolute nebula luminosity estimates, while the SMC itself is sufficiently small in angular size ($\sim320\arcmin\times185$\arcmin) that it can be easily studied in its entirety with current wide-field survey CCD cameras with modest numbers of exposures.

In this pilot paper we present discovery of our first 4 spectroscopically confirmed SMC PNe uncovered from a uniform, arcsecond resolution, deep, $\sim$5-Rayleigh (R$_{equiv}\sim$21) \ha{}  and matching SR-band map of the SMC. A key motivating factor in uncovering significant new SMC PNe is in the provision of the most complete SMC planetary nebulae luminosity function (PNLF) possible, akin to that produced in the LMC by \citet{2010MNRAS.405.1349R}. The exponential fit to the PNLF bright end cut-off has been shown to provide an absolute magnitude fiducial that is a powerful standard candle, on a par with the best cosmological distance indicators such as Cepheid variable stars, but can probe further out to the Virgo and Coma galaxy clusters and beyond \citep{2002ApJ...577...31C, 2005ApJ...629..499C, 2010PASA...27..149C}. It also provides clues to the underlying nature of the host galaxy's entire PN population as it represents an effective co-eval snapshot of all progenitor stars currently going through this brief evolutionary phase and the dominant different mass and age populations that are present and now being revealed \citep{2015ApJ...804L..25B}. Our anticipated expanded list of known PNe in the SMC, by up to 50\%, will permit more extreme ends of the PNLF to be explored,  provide a more complete SMC PNe sample for detailed study that includes lower-surface brightness, more evolved and previously under represented PN evolutionary stages.

This paper is structured as follows. In section 2  we provide some basic background on the known SMC PNe population, in section 3 we outline the multi-wavelength processes we have adopted to uncover new SMC PNe candidates from our data and vett previous PNe compilations. In section 4 we present the spectroscopic confirmation and analysis of our  preliminary sample while some brief discussion and our conclusions are given in section 5.

\section{The Magellanic Clouds and previous SMC PN surveys}

The Magellanic Clouds are part of a much larger system, including the Magellanic Bridge, the Interface Region, the Magellanic Stream and the Leading Arm \citep{2005A&A...432...45B}. Both galaxies are classified as dwarf irregulars, with well determined distances: $\sim$50 $\pm$3 kpc for the LMC \citep{1998ApJ...492..110M, 2000ApJ...529..786M, 2006MNRAS.365..401R} and $\sim$61$\pm$1 kpc for the SMC \citep{2005MNRAS.357..304H}.

The Magellanic Clouds contain a large reservoir of gas and their gas to dust ratios are higher by factors of $\sim$4 in the LMC and $\sim$17 in the SMC compared to our own 
Galaxy \citep{1984IAUS..108..333K}. They are rich in a variety of both stellar, compact and resolved emission-line sources. These include Luminous blue variables (LBVs), symbiotic stars, Cataclysmic variables (CVs), Wolf-Rayet stars (and shells), Supernova remnants (SNRs), super bubbles,  compact H\textsc{ii} regions and PNe. The number of known LMC PNe  has been significantly increased in recent years from $\sim$300 to $\sim$900 largely due to the major discoveries made by \citet{2006MNRAS.365..401R, 2006MNRAS.373..521R} based on very similar detection techniques to those presented here. However, the number of known PNe in the SMC has so far remained modest ($\sim$100). Moreover, the current sample consists of quite a heterogeneous compilation obtained from surveys which differ in depth, selection technique, resolution and areal coverage. This has limited their use for representative SMC PNe population studies and associated and accurate estimation of key PN parameters for a statistically significant SMC PNe sample. It is crucial to meaningfully study their evolution, the stellar evolutionary history of the SMC, the enrichment of the interstellar medium and the mass-loss history of their CSPN. These factors have helped motivate this current project.

\subsection{SMC PN surveys}

The earliest work on the SMC PN population was undertaken with a variety of modest aperture, wide-field, Schmidt-type telescopes fitted with objective prism dispersers 
\citep{1956ApJS....2..315H,1961AJ.....66..169L}. The first major study \citep{1956ApJS....2..315H} published the positions of 236 emission-line stars and 532 emission nebulae in both 
Magellanic Clouds. Twenty of the nebulae in the SMC were listed as PNe while nine were considered to be good PN candidates. \cite{1961AJ.....66..169L} produced a catalogue of 593 
'unresolved' emission-line objects in the SMC, of which 26 are considered to be PNe and 13 are probable PNe.

\cite{1978PASP...90..621S} later published another catalogue, based on a collection of various types of deep objective-prism photographic plates, listing 25 sources as PNe and only three as PN candidates. \cite{1980ApJS...42....1J} reported 8 known PNe and 19 candidates, while the work of \cite{1981PASP...93..431S} added 6 more candidates. Low dispersion objective prism spectra are indicative at best and a significant number of objects from these catalogues have never undergone proper higher-resolution spectroscopic confirmation. Sometimes, the presence of emission lines other than \ha{} was enough for a source to be previously classified as a PN \citep{1961AJ.....66..169L}. In many cases the presence of a continuum together with the \ha{} line was sufficient for the object to be classified as an emission-line star rather than a PN. Furthermore, a number of objects have only been observed once at low resolution \citep{1995AAS112..445M} without important diagnostic line ratios being resolved. 

Most recently, \cite{2002AJ....123..269J} using the ESO 2.2m telescope with the 8K$\times$8K mosaic CCD camera, surveyed $2.8 deg^{2}$ of the central SMC region searching for faint PNe using on and off-band [OIII] imaging and follow-up confirmatory spectra of candidates with the CTIO 4m telescope. They reported 59 confirmed PNe including 25 new discoveries which still represent the single, largest increase in SMC PNe till now. They note a high incidence of new, faint PNe exhibiting strong \NII{}/\ha{} ratios compared to that seen in the previously known `brighter' SMC PNe population. They attribute this to a selection bias to favour chemically enriched Type~I PNe \citep{1994MNRAS.271..257K} from higher mass and therefore younger progenitors being both shorter lived and partially self-obscured by dust and so harder to detect.

\section{A New Population of SMC PNe}

SMC PNe are an effectively co-located population in a coherent system at known distance. As \ha{} emission alone cannot distinguish between many emission-
object types and considering  \OIII{} emission is weak or even absent in some low excitation PNe  or those that are heavily obscured \citep{2010PASA...27..129F}, a different, systematic 
approach is required. This is both to confirm the veracity of many of the previously identified SMC PNe currently lacking unambiguous spectroscopic confirmation and to uncover new 
candidates. We have adopted a multi-wavelength approach. First we used a deep, narrow-band, arc second resolution \ha{} and matching broad-band ``SR'' survey of the whole SMC  to identify primary candidates. This map was created by \scos{} \citep{2001MNRAS.326.1279H} scans of original, high-quality Technical-pan photographic \ha{} imagery from the UKST (see Parker et al. \citeyear{2005MNRAS.362..689P} for general details of the UKST \ha{} surveys and Parker \& Bland-Hawthorn \citeyear{1998PASA...15...33P} for details of the narrow-band filter).

We then combined this with other available optical and mid-infrared (MIR) data to detect both low-excitation PNe candidates, extend PNe searches to fainter limits and to identify likely contaminants. Our broader goals are to provide a more spatially complete and refined catalogue of PNe with mimics removed and to identify many other emission objects (EmOs) across the SMC.

\subsection{Re-evaluation of the known SMC population}

To create a more complete and reliable inventory of the currently known SMC PNe population, we first evaluated the content of all previously published SMC emission-line catalogues, 
including those that include EmOs in general. This was due to the possibility of uncovering PNe mis-classified as some other emission object-type. At the SMC distance most PNe are very compact/barely resolved and there has been confusion in the literature about the identities of many SMC emission line sources. Application of our enhanced candidate identification and mimic elimination techniques based on the work of Frew \& Parker \citeyear{2010PASA...27..129F} together with the availability of new optical and multi-wavelength SMC imaging affords fresh opportunities for critical re-evaluation. The ten most comprehensive SMC emission object catalogues available in the literature and selected for review are listed in Table \ref{tbl:Catalogues}. They are of varying quality and integrity in terms of positional accuracy and identification and with samples that strongly overlap between catalogues so a critical re-evaluation is definitely worthwhile.

\begin{table*}
\begin{minipage}{124mm}
\footnotesize
\begin{center}
\caption{Principal SMC PNe catalogues with adopted abbreviations. The column headed `Known' shows the number of known PNe retrieved in a catalogue, the column `New' the number of new PN candidates, the column `Total' the total number of PNe and the column `Area covered' the areal coverage of the given survey.}
\begin{tabular}{llcccc}
\hline
PNe Catalogue	 & Abb. & Known & New & Total & Area covered\\
\hline
\cite{1956ApJS....2..315H} & HEN & 20 & 9 & 29 & 15 deg$^2$ \\
\cite{1961AJ.....66..169L} & LIN & 26 & 13 & 39 & $\sim$12 deg$^2$ \\
\cite{1978PASP...90..621S} & SMP & 25 & 3 & 28 & 30 deg$^2$\\
\cite{1980ApJS...42....1J} & J & 8 & 19 & 27 & $\sim$15 deg$^2$\\
\cite{1981PASP...93..431S} & SP & 28 & 6 & 34 & $\sim$20 deg$^2$\\
\cite{1985MNRAS.213..491M} & MG & 34 & 10 & 44 & 36 deg$^2$\\
\cite{1993AAS...102..451M}$^*$ & MA & 49 & 13 & 62 & 3.4 deg$^2$\\
\cite{1995AAS112..445M}  & M & 53 & 9 & 62 & $\sim$ 15 deg$^2$\\
\cite{2000MNRAS.311..741M}${^*}{^*}$ & MB & 23 & 108 & 131 & 49 deg$^2$\\
\cite{2002AJ....123..269J} & JD & 34 & 25 & 59 & 2.8 deg$^2$\\
\hline
Total number of current non-overlapping SMC PNe: &  & 96 &  &  & $\sim$120 deg$^2$ \\
\hline 
\end{tabular}
\label{tbl:Catalogues}
\end{center}
*Catalogue includes PNe and VLE objects\\
* *Differences in numbers of newly identified PN candidates are due to different selection technique and larger survey area.
\end{minipage}
\end{table*}

All catalogues of existing EmOs were cross checked against each other by their reported coordinates and previous classifications, and then referred back to our new \ha{} SMC image 
map. If two sources had similar but slightly different coordinates they were considered to be the same object if only one clear emission source was evident in our own \ha{} data at or near the reported location. Conversely if a catalogued source had no apparent emission equivalent in our data or related imagery, e.g. from the Magellanic Cloud Emission-Line Survey (MCELS) \citet{2005AAS...207.2507S}, then its identity as a true emission-line source is called into question. Detailed results from this evaluative study will be presented in a later paper. An astrometric WCS grid, accurate to $\sim0.2\arcsec{}$, has been carefully applied to our digital \ha{} and `SR' SMC maps following standard \scos{} procedures, e.g. \citet{2001MNRAS.326.1279H,2001MNRAS.326.1315H}. This coordinate system provides the basis for all new and previously identified PNe presented in this and our subsequent papers.

As a result of this careful cross-checking, a unique list of 247 objects, including PNe, emission line stars, red giant branch stars, asymptotic giant branch stars, spectroscopic binaries and other miscellaneous EmOs were compiled. These objects were then divided in two groups. The first comprising 96 PNe (either previously confirmed or previously identified as a PN candidate) and the second 151 miscellaneous EmOs, such as young stellar objects, various types of emission line stars, H\textsc{ii} regions, SNRs, etc. This was based primarily on whether they have spectroscopic confirmation, either when they were first identified or in subsequent investigations, and secondly on the published likely identification and/or evidence supplied to support a given classification.

\subsection{Discovery Technique}

We have carefully examined the UKST \ha{ }on-band and contemporaneous broad-band red continuum `SR' digital image data for over 120 $\deg^2$ centred in and around the SMC. As a result we have uncovered a new population of over 50 SMC PNe candidates while also calling into doubt the veracity of some previously reported SMC PNe. Our search process adopted the same powerful and proven colour merging technique successfully applied to the LMC by \citet{2006MNRAS.365..401R, 2006MNRAS.373..521R, 2013MNRAS.436..604R}. Using the \textsc{KARMA} visualisation software package, we surveyed ten fields, each $\sim4\arcdeg\times3\arcdeg$ in size, covering the main body and the outskirts of the galaxy. We created merged false-colour images of all ten fields, combining \ha{ }(coloured in blue) and broad-band (coloured in red) images. The images were then overlaid with annotation files containing the positions of all previously known 247 cross-checked emission objects of all kinds. This enabled us to easily determine the merged colour appearance of true PNe and other types of EmOs.

All ten fields were systematically searched by visual scanning, looking for a faint, compact or barely resolved objects. 
By carefully choosing the image combination parameters we could perfectly balance the intensity of \ha{} and broad-band red matching images. This allows only specific features of 
one or other pass-band to be observed, because the long 3 hour \ha{} exposures and short 15 minutes broad-band SR exposures are well matched to depth for continuum point sources. 
In our chosen pass-band merging scheme emission objects such as PNe and \HII{} regions appear with a uniform strong blue colour, whereas normal continuum stars are a uniform pink-purple colour. Emission-line stars usually have a strong continuum component and so are easily detected by the narrow extent of their blue halos around a strong pink core 
\citep{2006MNRAS.365..401R}.

The RA/DEC positions of all candidates were recorded in J2000 coordinates based on the accurate WCS in our data and cross-checked for entries in SIMBAD\footnote{http://simbad.u-strasbg.fr/simbad.} using a search radius of $1\arcmin$ (due to poor astrometric accuracy of early listings). If no matching astronomical object was found, the source was confirmed as a new candidate emission source. This careful scanning of our colour-merged  \ha{}/SR imagery provided an initial sample of 67 PNe candidates. Many other new emission sources were also uncovered during this process but they are not included in this sample as they are either clearly stellar or have angular (and hence physical) extents that are large enough to rule out a PNe identification.

\subsection{Multi-wavelength Images and cross-identification}

PNe emission candidates uncovered by careful scrutiny of our false-colour \ha{}/SR imagery were cross-checked against multi-wavelength data obtained from a range of other
available optical, near-infrared (NIR), mid-infrared (MIR) and radio surveys, as listed in Table \ref{tbl:MultiwavelengthSurveys}, to see if they were also detected in these data. If a PN 
candidate exhibited evidence of emission line flux in at least two of our combined colour image renditions in the optical range then it was deemed to be a corroborated candidate. These comprised our false colour \ha{}/SR merged images, the quotient image (\ha{} divided by SR), the \scos{} broad-band `SSS' blue $B_{j}$, red R and NIR 'I'-band combination or the lower resolution MCELS emission line narrow-band combined colour image. MCELS is a deep multiple emission-line CCD survey of both Magellanic Clouds, with images in \ha{}, \SII{} and \OIII{} \citep{2005AAS...207.2507S}. In order to allow subtraction of the stellar background and leave only emission-line objects, two MCELS continuum-band images are available. 
Many of our candidates were independently detected in the MCELS data but due to the faintness of many of our new candidates they are generally too faint to be seen in the SSS. 
Some candidates are only detected in our \ha{}/SR imagery but this does not rule them out as being real objects.

Our PN candidates were also cross-checked for counterparts against the Two Micron All Sky Survey (2MASS) NIR image data and the sensitive but lower resolution Wide-field Infrared 
Survey Explorer (WISE) MIR imagery, \citet{2010AJ....140.1868W} at 3.4, 4.6, 12 and 22$\mu$m (bands named W1 to W4) and the Molonglo Galactic Plane Survey 2nd Epoch 
(MGPS2) 843 MHz radio images. These data can provide additional diagnostic power in cases where multiple detections are made e.g. \citet{2012MNRAS.427.3016P}, allowing many mimics in previous compilations to be identified.

PNe tend to have a narrow-range of distinctive appearance in false-colour images from different surveys that can assist identification. For example, \citet{2007ApJ...669..343C} have shown that the dominant colours of PNe in their study of \textit{Spitzer} MIR imagery are red, orange and violet when combining the different MIR bands as false-colour RGB images. The dominant colours of PNe in \scos{} Sky Survey (SSS)  RGB images are green or turquoise due to the significant contribution from the strong \OIII{} PN emission line and/or the \ha{}/\NII{} lines in the red band. 

Following this careful cross-check we narrowed the list to 50 high-quality PN candidates by excluding objects that did not appear in any of these other surveys and that also had other image characteristics that indicated they were possible artefacts or plate flaws in the \ha{}/SR data. We selected four candidates for initial spectroscopic follow-up during a brief window of opportunity in another of our observing runs. The effectiveness of our multi-wavelength selection technique to deliver high quality PN candidates was shown by the subsequent spectroscopic PN confirmation of all four preliminary targets (100\% success). This is perhaps unsurprising given the equivalent high success rate for PN confirmation in the LMC by \citet{2006MNRAS.365..401R, 2006MNRAS.373..521R} who used almost identical selection technique. Figure \ref{fig:Objects} displays false colour \ha{}/SR-band and quotient discovery images of these candidates, together with the SSS  I, R and $B_{j}$ combined RGB images. Only the first and brightest of our new PNe, DPR1, has a clear SSS counterpart though very faint detections of the others are seen. This source is also just visible in MCELS but not at the level were its emission nature would have been obvious. The adopted naming scheme for our new PNe follows our usual process of using the surname initial of the key people involved in the project - here the first three authors (PhD student and two supervisors).

\begin{table*}
\begin{minipage}{124mm}
\footnotesize
\begin{center}
\caption{Multi-wavelength sky surveys cross-checked.}
\begin{tabular}{lcccc}
\hline
Sky Survey & Wavelength & Filters & $\lambda_{c}$ (\AA) & $\Delta\lambda$ (\AA)\\
\hline
Optical & &\\
\hspace{0.5cm}MCELS \citep{2005AAS...207.2507S} & $(498\mbox{--}689)\,\mbox{nm}$ & \ha{} & 6563 & 30\\
 										 &                                                                   & \SII{} & 6742 & 50\\
										 &							     & \OIII{} & 5007 & 40\\
										 &							     & continuum-band & 6850 & 95\\
										 &							     & continuum-band & 5130 & 155\\	
\hspace{0.5cm}SHS \citep{2005MNRAS.362..689P} & $(590\mbox{--}690)\,\mbox{nm}$ & \ha{} + \NII{} & 6590 & 75\\
										&						     & SR & 6445 & 1000\\	
Optical and Near-Infrared & &\\
\hspace{0.5cm}SSS \citep{2001MNRAS.326.1279H} & $(400\mbox{--}900)\,\mbox{nm}$ & B$_{j}$ & 4450 & 94\\
										&						      & R & 6580 & 1380\\
										&						      & I & 8060 & 149\\
\hline
Sky Survey & Wavelength & Filters & $\lambda_{c} (\mu{}m$) & $\Delta\lambda (\mu{}m$)\\																		\hline	      		
Near-Infrared & &\\
\hspace{0.5cm}2MASS \citep{2006AJ....131.1163S} & $(1.25\mbox{--}2.16)\,\mu\mbox{m}$ & J & 1.235 & 0.162\\
									        &							 & H & 1.662 & 0.251\\
									        &						          & K$_{s}$ & 2.159 & 0.262\\
Mid-Infrared & &\\
\hspace{0.5cm}WISE \citep{2010AJ....140.1868W} & $(3.4\mbox{--}22)\,\mu\mbox{m}$ & W1 & 3.35 & 0.66 \\
									      &							   & W2 & 4.60 & 1.04 \\
									      &							   & W3 & 11.56 & 5.51 \\
									      &							   & W4 & 22.09 & 4.10 \\
Radio & &\\
\hspace{0.5cm}MGPS2 \citep{2002IAUS..199..259G} & $\sim35.6\,\mbox{cm}$ &0 & / & / \\
\hline
\end{tabular}
\label{tbl:MultiwavelengthSurveys}
\end{center}
\end{minipage}
\end{table*}

%Figure1-DPR1, DPR2, DPR3 and DPR4

 \begin{figure*}
 \centering
 
  \includegraphics[scale=0.8]{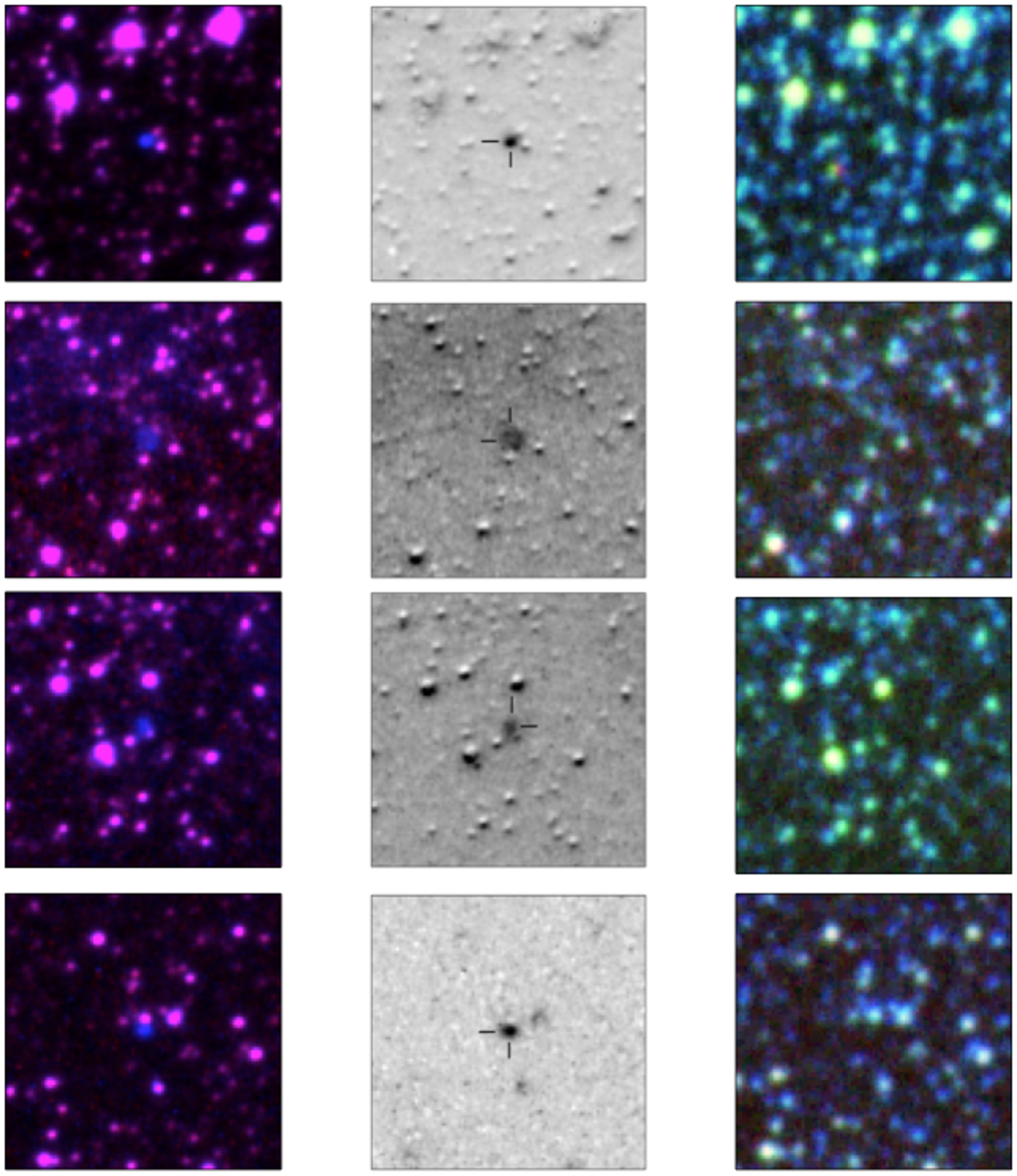}
   
   \caption{Image types from left to right are: \ha{}/SR merged false-colour, \ha{}/SR-quotient and SSS (IR$B_j$), with DPR1 to DPR4 from top to bottom. 
  Image sizes are $1.1\arcmin\times1.1\arcmin$ with North-East to the top left.}
 \label{fig:Objects}
 \end{figure*}

\section{Spectroscopic Confirmation}

Four PNe candidates from our new compilation named DPR1 to DPR4 were observed by one of us (QAP) using the Cassegrain spectrograph on the South African Astronomical 
Observatory's (SAAO) $1.9\,\mbox{m}$ Radcliffe telescope over a 5 night period in March 2014 in order to test the efficacy of our selection process. The standard slit-spectrograph was used with a SITe $266\times1798$ pixel CCD giving 0.5$\arcsec$/pixel on the slit with a slit scale of $6\arcsec$/mm. It was used with low resolution grating `7' (300 lines/mm with 5\AA{} resolution and dispersion of 210\AA{}/mm) at an angle of $17\arcdeg24\arcmin$. This provides a broad spectral range from 3500-7400\AA{} which covers all  the strongest PN optical emission lines for a dispersion of 1.8 \AA{}/pixel.

Appropriate bias frames, arc calibration exposures, dome and twilight flat fields and radial velocity and spectrophotometric standard stars were taken to assist with the standard spectral reduction, flux calibration and quality control process. For DPR1, DPR2 and DPR4 the slit width was $2\arcsec$. For the more resolved candidate DPR3 a $3\arcsec$ slit was used to capture more of the object's flux at the expense of a modest reduction in spectroscopic resolution due to the poorer seeing during this exposure. For all targets, the same spectrophotometric standard star LTT3864 was observed.

Exposures were carefully reduced with the Image Reduction and Analysis Facility (IRAF) V2.16\footnote{IRAF is distributed by the National Optical Astronomy Observatory, which is operated by the Association of Universities for Research in Astronomy (AURA) under a cooperative agreement with the National Science Foundation.} \citep{1986SPIE..627..733T, 1993ASPC...52..173T} following standard procedures. These steps include frame trimming, average bias frame subtraction, dome and twilight flat-fielding, cosmic ray removal, 1-D spectral extraction, wavelength calibration, sky-subtraction and flux calibration. The observing conditions were generally stable and the slit width was adjusted according to the seeing conditions and in the case of spectrophotometric standard star LTT3864,  widened to $5\arcsec$ to ensure all flux was collected.

Each observed candidate was confirmed as an emission line source and exhibited no obvious stellar continuum. This eliminates confusion with any emission line star for the two compact sources DPR1 and DPR4. Each candidate also gave high ratios of \NII{} to \ha{} that are not observed in \HII{} regions, which are the most likely narrow-emission line contaminant \citep{2000ApJ...537..589K}. Unfortunately, none of the spectra appear to exhibit the \OIII{} or \hb{} emission lines in the blue. This indicates the possible effects of some modest extinction but more seriously the limited blue S/N obtained for these faint sources, especially for the more clearly resolved and lower surface-brightness candidates DPR2 and DPR3. 

The applied flux calibration is indicative only but measured line ratios should be quite reliable in principle due to the close wavelength proximity of the lines used, modulo the effects of low S/N for DPR2 and DPR3. A summary of observed line ratios (and errors), radial velocities and estimated angular and physical size for these new PN candidates is given in Table \ref{tbl:CandidateDetails}. A discussion of the results of each observed candidate are given below.

%Table 3
\begin{table*}
\footnotesize
\begin{center}
\caption{Observing log of spectroscopically confirmed new PNe candidates towards the SMC with observed spectroscopic relative line intensity $(F)$ ratios, their radial velocities, angular and physical sizes.}
\begin{tabular}{lcccccccccccc}
\hline
Name & RA       & Dec                                               & Date  & Time & $T_{\mbox{exp}}$ & F \NII{}/F \ha{}  & $<V_{r}>$  & $\sigma_{n}$ & $\Theta (\arcsec)$ &  $d$ & Seeing\\
	  & (h m s) & ($\arcdeg \: \arcmin \: \arcsec$) &            & (UT) &          (s)                   &                            & $(km/s)$	 & $(km/s)$         & upper limit               &   (pc)            & (\arcsec)\\
\hline
DPR1 & $00\:54\:26.00$ & $-72\:30\:59.4$ & 7 Mar 2014 & 17:52:42 & 1200 & 7.0$\pm$0.64 & 80 & 10 & unresolved & / & 2.2\\
DPR2 & $00\:59\:47.60$ & $-72\:32\:31.4$ & 11 Mar 2014 & 18:01:46 & 1130 & 2.2$\pm$0.80 & 169 & 4.5 & 7.5 & 2.2 & 2.9\\
DPR3 & $01\:02\:55.20$ & $-72\:21\:32.0$ & 9 Mar 2014 & 18:00:10 & 1800 & 3.7$\pm$1.00 & 241 & 16 & 4.5 & 1.3 & 2.4\\
DPR4 & $01\:04\:30.00$ & $-73\:15\:10.5$ & 8 Mar 2014 & 17:59:24 & 1200 & 7.7$\pm$0.75 & 167 & 18.5 & unresolved & / & 2.5\\
\hline
\end{tabular}
\label{tbl:CandidateDetails}
\end{center}
\end{table*}

\subsection{DPR1}

The spectrum of this compact, unresolved PN candidate DPR1 is of reasonable S/N and exhibits no stellar continuum and only narrow-lined \ha{ }$\lambda6563$\AA{} emission and very strong \NII{ }$\lambda\lambda6548,6584$\AA{} lines. There are no evident \OIII{} or \hb{} lines in the blue or clear detection of [SII] in the red. The red emission lines effectively rule out confusion with any emission line star or compact \HII{} region. For more accurate line ratio determination and possible detection of fainter emission lines we intend to re-observe all four candidates on a larger telescope. 

The sky spectrum was well determined for this object because the angular size of the target was much smaller than the length of the spectrograph slit which provides ample adjacent sky regions. This is seen by the lack any residual  presence of the very strong \OI{} night sky line at $5577$\AA{} and also the lack of residuals from the weaker \OI{} $6363$\AA{} line and only a small residual from the stronger $6300$\AA{}  line in the red. Hence, one can be confident that the observed, low \ha{} emission is real and not due to an over subtracted \ha{} sky component. In such low resolution spectra the observed auroral component at effectively zero velocity and that expected for a PN in the SMC of $\sim$160 km/s will not present much of an offset. 

Furthermore, any Galactic \ha{} emission component would be very weak at this Galactic latitude ($\sim-44$ degrees) while an examination of Figure \ref{fig:Objects} indicates absence of any local diffuse SMC emission component. Deblended Gaussian fits were applied to the closely spaced \NII{} and \ha{} emission lines to provide radial velocity estimates and relative line fluxes for measuring the line ratios. A large velocity gradient for different object types is observed in the SMC, ranging from 90 km/s to 200 km/s, as shown by \citet{1987MNRAS.226..513T} and \citet{ 2004ApJ...604..176S}. The DPR1 lines provide an average radial velocity of $<V_{r}>$ = 80 km/s, $\sigma_{n}$= 10 km/s and $n=3$ (including heliocentric velocity correction) which is little low for membership of the SMC. On the basis of our combined imaging and new spectroscopic evidence we identify DPR1 as a  likely Type~I PNe , largely due to the very high \NII/\ha{} ratio of 7.0 and weak/absent [SII].  Without measurable \OIII{} lines in the spectrum this object likely falls into the low excitation category but the absence of detectable \hb{} also indicates the effects of poor S/N in the blue.

\begin{figure*}
 \centering
 \includegraphics[scale=0.5,angle=270]{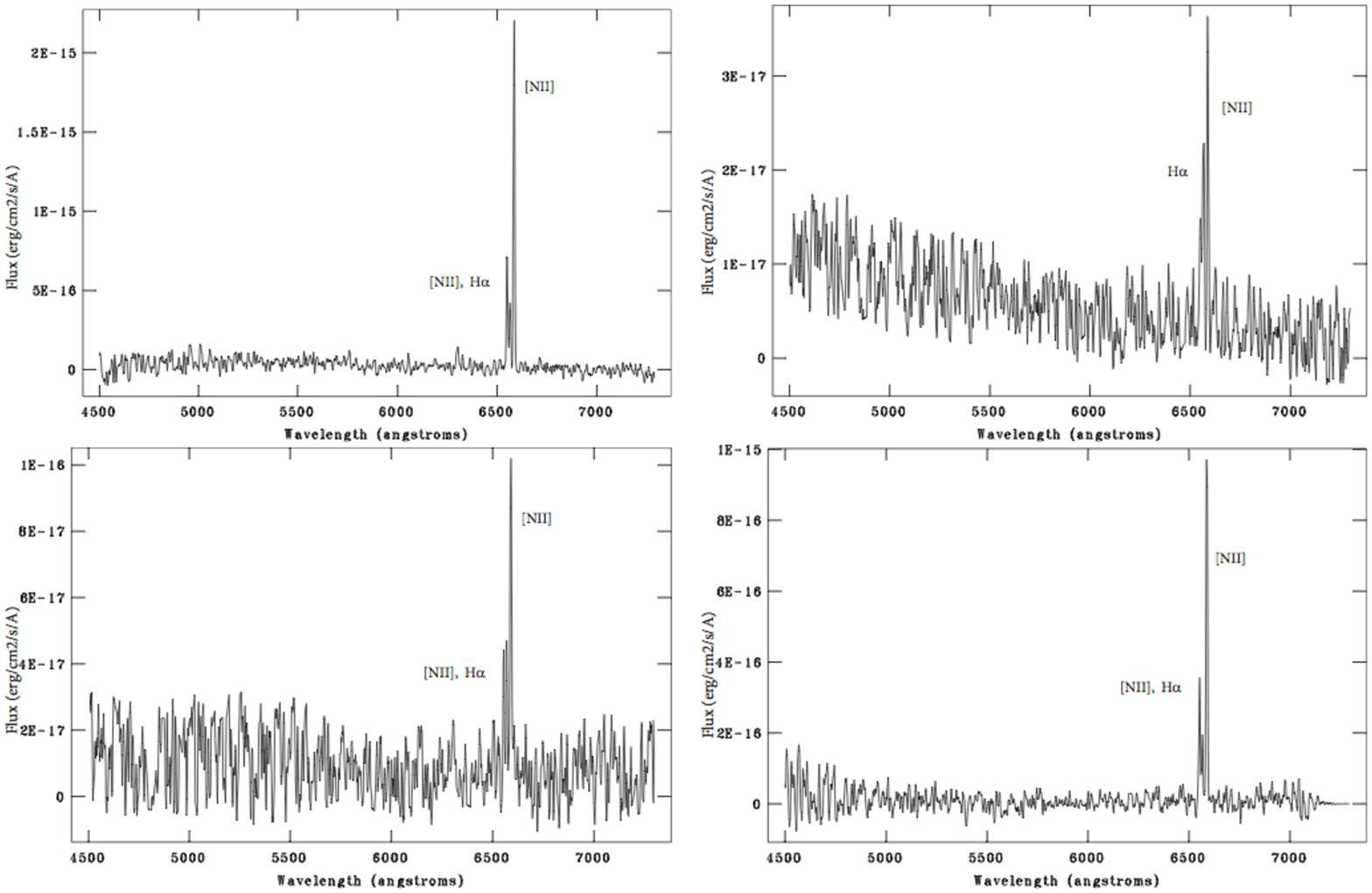}

\caption{SAAO spectra of newly discovered PNe in the SMC. \textit{Top left:} low excitation nebula DPR1 with observed high \NII{}/\ha{} ratio. Note the absence of \OIII{} emission lines. 
\textit{Top right:} low signal-to-noise ratio (S/N) spectra of low excitation nebula DPR2. Some artefacts of airglow processing remain, but nebula \NII{} and \ha{} emission are discernible above the general background noise. \textit{Bottom left:} low (S/N) spectra of low excitation nebula DPR3. Airglow processing artefacts again remain, but nebula \ha{} and \NII{} emission is discernible. \textit{Bottom right:} low excitation nebula DPR4, again with a high \NII{}/\ha{} ratio, similar to DPR1. In common with other candidates, \OIII{} lines are not detected.}
\label{fig:DPRSpectra}
\end{figure*}

\subsection{DPR2 and DPR3}

These two low-surface brightness candidates are clearly resolved as can be seen in their quotient images in Figure \ref{fig:Objects}, with major axes of 7.5 and 4.5 arc seconds respectively (determined from FWHM of the brightness profile), eliminating any possible confusion with emission line stars. At the nominal distance of the SMC this corresponds to maximum physical sizes of 2.2 and 1.3 parsecs (with seeing estimated at 2.5\arcsec{}) which are at the upper size limit for Galactic PN. For example, \citet{2006MNRAS.372.1081F} reported discovery of two highly evolved Galactic PNe, RCW24 and RCW69, with physical dimensions larger than 5 pc and 1.3 pc respectively, while \citet{2004PASA...21..334P} reported PFP1 with diameter of $\sim$1.5 pc. 

Confusion with young SNRs is also ruled out by the absence of any strong \SII{} and hence shocked conditions. Their observed spectra, although noisy, are also very similar. Both have high \NII{}/\ha{} ratios of between 2 and 3 - still well in excess of that exhibited by \HII{} regions. A PN identification for both sources is consequently strongly favoured. Gaussian line de-blending was carefully applied to the closely spaced \NII{}, \ha{} emission lines to provide radial velocity estimates and relative line fluxes for measuring the line ratios which in these two cases have larger associated errors. The measured radial velocity for DPR2 is $<V_{r}>$ = 169 km/s, $\sigma_{n}$= 4.5 km/s, $n=3$ and for DPR3 is $<V_{r}>$ = 241 km/s, $
\sigma_{n}$= 16 km/s, $n=3$, both with heliocentric velocity correction applied. The result for DPR2 is in excellent agreement with the SMC velocity within the errors while that for DPR3 
is about 50 km/s high.

Artefacts from night sky-subtraction due to low S/N are present, with contributions from \OI{ }$\lambda\lambda5577,6300,6364$\AA. These are removed from the spectra for these two cases so the PN emission lines are easier to see. Improved S/N on a large aperture telescope is needed to provide other plasma diagnostics as well as the ability to work out extinction directly from the Balmer decrement. The imaging and spectroscopic evidence collected nonetheless strongly support assessment of DPR2 and DPR3 as PNe. They are also likely highly evolved, low-excitation Type~I PNe given the observed  \NII{}/\ha{} ratios and sizes but the poor S/N prevents detection of \hb{} or any  \OIII{} lines.

\subsection{DPR4}

The spectrum of our final candidate DPR4, another compact object, has reasonable S/N and exhibits strong \NII{} emission similar to DPR1. Again there is the absence of any obvious \OIII{} or \hb{} lines in the blue likely due to low S/N given the strength of \hb{}. Only \ha{} and \NII{}$\lambda \lambda 6548,6584$\AA{} lines are present, with the observed \NII{}/\ha{} ratio of 7.7, the highest of all four candidates. Confusion with compact \HII{} regions or emission lines stars are ruled out by the observed narrow lines ratios and the absence of any stellar continuum. Again sky subtraction is excellent with no residual \OI{} at $5577,6300, 6363$\AA{} being evident and this process does not contribute to any over-subtraction of \ha{}, while Figure~1 also shows an absence of any local SMC diffuse \ha{} emission component. 

Deblended Gaussian fits were applied to the \NII{}, \ha{}, \NII{} lines to permit radial velocity estimation and to provide the relative fluxes for the line ratio measurement. The average radial velocity obtained from these lines, including application of the appropriate heliocentric velocity correction, was $<V_{r}>$ = 167 km/s, $\sigma$= 18.5 km/s in excellent agreement with the canonical SMC radial velocity. As for previous candidates, re-observation of DPR4 is recommended under better observing conditions and on a larger telescope. Our imaging and spectroscopic data strongly support assessment of this candidate as another low excitation likely Type~I PN.

\section{Discussion \& Conclusions}

We have undertaken a survey for faint PNe in the SMC based on careful examination of deep, arcsecond resolution, UKST \ha{} narrow-band and equivalent red band `SR' imaging data. Candidates were selected by a systematic visual search of ten SMC fields covering the entire body and the outskirts of the galaxy over an $\sim14\arcdeg \times19\arcdeg$ area. The discovery process used the effective, proven colour merging technique described in detail in section 3.2 and used successfully for the LMC by \citet{2006MNRAS.365..401R, 2006MNRAS.373..521R, 2013MNRAS.436..604R}.  False colour MCELS and SSS RGB images were created for all candidates to complement the \ha{}/SR merged colour images and then used in cross-checking candidates against a range of other NIR, MIR and radio survey data (as detailed in Table \ref{tbl:MultiwavelengthSurveys}) looking for counterparts. These comparisons provided strong evidence of false classifications for many previously identified SMC PNe that currently lack decent confirmatory spectroscopy. These results will be presented in an associated paper.

The positions of all emission sources uncovered were also cross-checked against existing PNe catalogues and general SMC EmOs compilations for previous detections. This enabled the identification of 50 new PN candidates.  Four of these were selected for preliminary spectroscopic follow-up and all were subsequently confirmed as bona-fide PNe. The positions of all our newly discovered candidates are shown in Figure \ref{fig:SMCArea} which gives the SMC MCELS \ha{} mosaic image overlaid with the positions of the 96 currently known SMC PNe (spectroscopically confirmed and candidates) as red circles, our 46 new candidates as green circles and the 4 new PNe confirmed in this study as blue circles.

%Figure3
\begin{figure*}
  \centering
    \includegraphics[width=\textwidth]{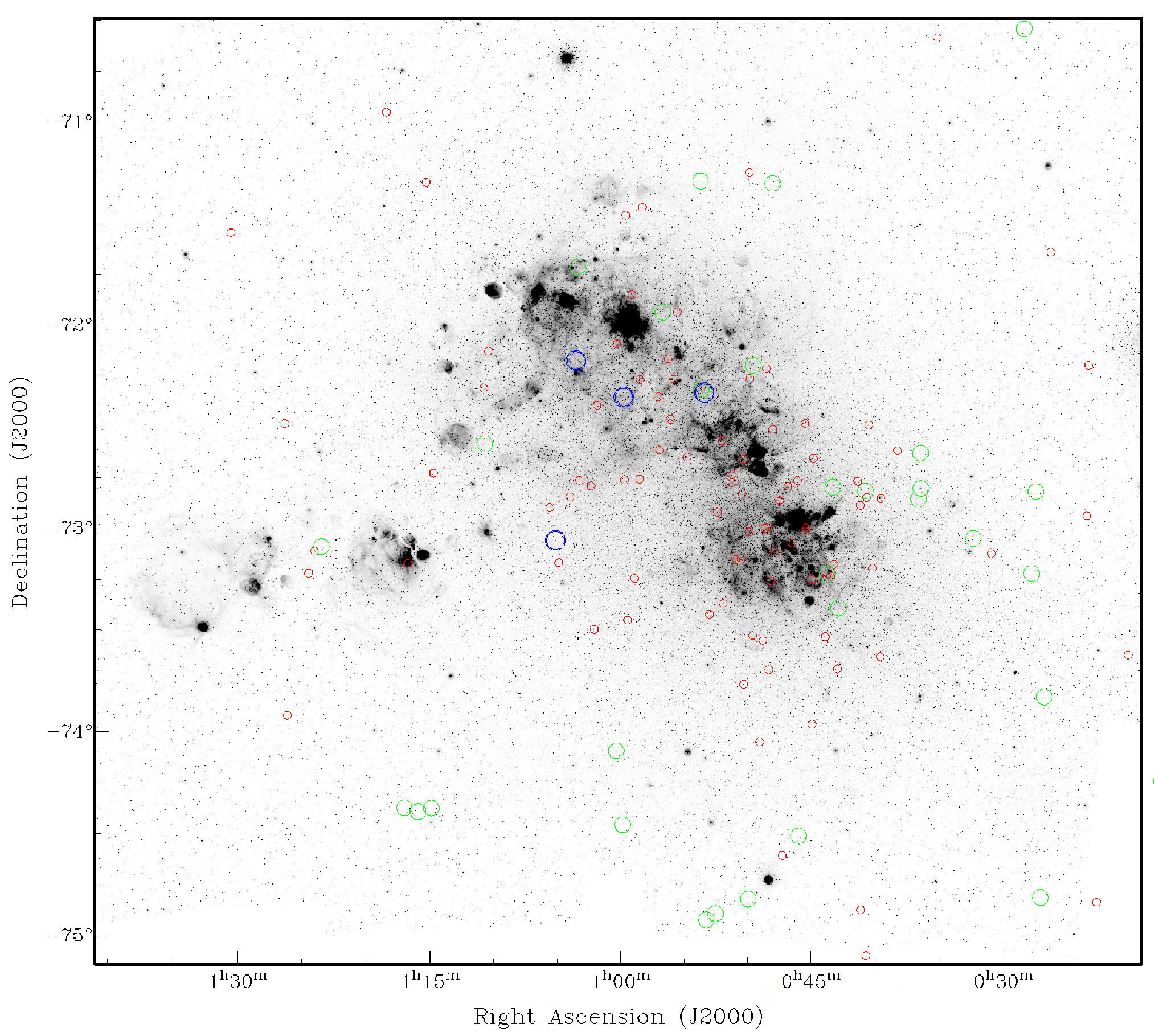}
  \caption{SMC MCELS mosaic \ha{} image with positions of 96 known PNe (confirmed and candidates) \textit{(small red circles)}, 46 new candidates \textit{(green circles)} and 4 new PNe confirmed in this study \textit{(large blue circles)}.}
  \label{fig:SMCArea}
    \end{figure*}

The confirmatory spectra for these candidates are all somewhat homogenous. Their observed \NII{}/\ha{} line ratios are all very high ranging from 2 to 7. This rules out any possible confusion with \HII{} regions which never show such ratios, while other spectral features exclude confusion with SNRs (absence of any \SII{} lines) or emission line stars (only narrow-lines and absence of any continuum). Galactic PNe similar to the new SMC PNe uncovered here include the evolved PNe IsWe\,1, RCW\,24 and WeDe\,1. These all exhibit 
strong \NII\ emission and weak to absent \OIII\ emission \citep{1986A&A...156..101G, 2006MNRAS.372.1081F} and low-luminosity central stars. These preliminary results demonstrate the strength and promise of our selection technique for providing high quality PNe candidates as well as demonstrating the importance of multi-wavelength comparison for refining identifications and unmasking mimics. 

Based on these preliminary results,  we expect to increase the number of SMC PNe by between 20 and 50\% after planned spectroscopic follow-up of all of our PN candidates and 
refinement of source identifications of existing SMC emission line calalogues. This work will include multi-wavelength analysis and spectroscopic confirmation (or rejection) of SMC emission-objects previously listed as PNe. Once complete, this survey will permit more extreme ends of the PNLF to be explored, especially at the faint end, and enable us to study under represented evolutionary stages of SMC PNe. It will provide a sample whose detailed study will have significant impact on improving abundances and rates of elemental enhancement in lower-mass stars within a low-metallicity environment. Other significant advances will come from the improved kinematical data for the galaxy derived from over 300 emission-line objects and their accurate radial velocities.

\section*{Acknowledgements}

We thank the anonymous referee who provided valuable and detailed comments which have significantly improved the quality of the paper. 
Financial support for this research was provided by Macquarie University Research Excellence Scholarship. Danica Dra\v{s}kovi\'{c} would also like to thank Macquarie 
University for a PhD scholarship. We thank South African Astronomical Observatory for observing time. We also thank Dr David Frew for useful discussions and Mr Travis Stenborg for help with the SAAO observations.

\bibliography{references}

\bsp

\label{lastpage}

\end{document}